\documentclass[showpacs,twocolumn,preprintnumbers,amsmath,amssymb]{revtex4}

\usepackage{graphicx}
\usepackage{dcolumn}
\usepackage{bm}

\newcommand{\bq}{\begin{equation}}
\newcommand{\eq}{\end{equation}}
\newcommand{\bqa}{\begin{eqnarray}}
\newcommand{\eqa}{\end{eqnarray}}

\def\be     {\begin{equation}}
\def\ee     {\end{equation}}
\def\bea        {\begin{eqnarray}}
\def\eea        {\end{eqnarray}}
\def\bnn    {\begin{eqnarray*}}
\def\enn    {\end{eqnarray*}}

\begin{document}

\title{Collective dynamics of Fermi-surface fluctuations in an interacting Weyl metal phase}
\author{Chungwon Jeong and Ki-Seok Kim}
\affiliation{Department of Physics, POSTECH, Pohang, Gyeongbuk 790-784, Korea}
\date{\today}

\begin{abstract}
Instabilities in a Landau's Fermi-liquid state occur, increasing the strength of interaction parameters in the Landau's Fermi-liquid theory. Introducing both the Berry curvature and chiral anomaly into this theoretical framework, we investigate collective dynamics of Fermi-surface fluctuations and reveal their instabilities in an interacting Weyl metal phase with broken time reversal symmetry. Recently, we proposed a topological Fermi-liquid theory to describe this interacting Weyl metal phase, where not only the Berry curvature but also the chiral anomaly is introduced into the Landau's Fermi-liquid theory [Phys. Rev. B \textbf{95}, 205113 (2017)]. Based on the Boltzmann-equation framework, we find criteria for the stability of the topological Fermi-liquid state as a function of forward scattering Landau's interaction parameters and the distance of a pair of Weyl points given by an external magnetic field. In addition to these instability criteria for general angular momentum channels, we investigate the dispersion relation of the zero-sound mode as the simplest example of such Fermi-surface fluctuations. Zero sound modes are well-defined collective excitations in a Landau's Fermi-liquid state, given by the collective dynamics of Fermi-surface deformations in the spin-singlet channel with zero angular momentum, where their instability is related with phase separation. We find that the role of the Berry curvature changes the instability criteria of the Landau's Fermi-liquid state. Even if the zero-sound mode is stable in the region of the forward-scattering amplitude, the Berry curvature gives rise to Landau damping beyond the Landau's Fermi-liquid theory. Based on the instability criterion of the zero-sound mode, we propose a phase diagram for a topological Fermi-liquid state against the phase separation in the plane of Landau's interaction parameter and effective Berry curvature, which generalizes the one-dimensional phase diagram of the Landau's Fermi-liquid theory.
\end{abstract}

\maketitle

\section{Introduction}

Electron correlations can give rise to fractionalized excitations in the presence of a topological structure \cite{Intro_FQHE}. It has been well established that a spin$-1/2$ chain shows gapless spectra, the nature of which is given by spin-fractionalized excitations, referred to as spinons and allowed by the topological $\theta = \pi$ term \cite{Spin_Chain_1_2}. This mathematical structure was generalized \cite{NLsM_Theta} and applied to a two dimensional spin$-1/2$ system with $SU(2)$ symmetry, where an emergent topological term may allow spinon excitations at a quantum critical point between an antiferromagnetic ordered state and a valence bond solid phase \cite{DQCP0,DQCP1,DQCP2,DQCP3}. This so called deconfined quantum critical point is forbidden in the Landau-Ginzburg theoretical framework.

It is natural to ask whether such fractionalized excitations can occur in quantum phase transitions of metals. However, symmetry breaking in a Landau's Fermi-liquid state is described by the Landau-Ginzburg effective field theory for local order parameter fields, which may not have quantum number fractionalization. Applying the mechanism discussed above into quantum phase transitions of metals, we need a topological term. Such a topological term, more precisely a topological-in-origin $\theta$ term \cite{Nielsen_Ninomiya,Burkov,TRB WM Anomaly,Anomaly_IS_KS}, appears in a Weyl metal phase \cite{WM1,WM2,WM3,WM4}, which can be realized by applying external magnetic fields into a spin-orbit coupled Dirac band structure and splitting the four-fold degeneracy into a chiral pair of two-fold degeneracy \cite{WM_Review1,WM_Review2,WM_Review3}. If symmetry breaking occurs in this Weyl metal state, an effective Landau-Ginzburg theory of the corresponding local order parameter field may have a topological-in-origin term, reflecting the chiral anomaly of the Weyl band structure and referred to as 't Hooft anomaly matching \cite{Anomaly_Matching}.

In the present study, we examine instabilities in a Weyl metal phase with broken time reversal symmetry as the first step for this research perspective. We consider collective dynamics of a pair of chiral Fermi surfaces. In the Landau's Fermi-liquid state, it is well known that such collective deformations of Fermi surfaces give rise to zero sound modes in the spin-singlet channel with zero angular momentum, well defined out of particle-hole continuum in the dispersion relation, where their instability is related with phase separation \cite{Negele_Orland,Nozieres_Pines}. In order to investigate the zero sound mode in an interacting Weyl metal phase, we need to generalize the Landau's Fermi-liquid theory, taking into account both the Berry curvature and chiral anomaly. Recently, we proposed a topological Fermi-liquid theory to describe an interacting Weyl metal phase with a pair of chiral Fermi surfaces in the absence of time reversal symmetry, where not only the topological information but also marginal forward scattering interactions are introduced \cite{TFL_Kim}. The term of ``topological Fermi-liquid theory" was originally coined by F. D. M. Haldane \cite{WM1}. We point out that Berry Fermi liquid theory has been developed based on the canonical quantization approach \cite{Berry_FL}, similar to the original proposal for the Landau's Fermi-liquid theory \cite{Negele_Orland}, while our approach is based on the path integral formulation, parallel with the Shankar's renormalization group analysis \cite{Shankar_RG_LFL}.

Following the Boltzmann transport theory in the Landau's Fermi-liquid state \cite{Negele_Orland,Nozieres_Pines}, we investigate collective dynamics of Fermi-surface fluctuations and study their instabilities as a function of the external magnetic field and the forward scattering amplitude in a topological Fermi-liquid phase \cite{TFL_Kim,Berry_FL}. In addition to these stability criteria of the topological Fermi-liquid phase for general angular momentum channels, we examine the dispersion relation of the zero sound mode as the simplest dynamics of the pair of chiral Fermi surfaces. We find that the role of the Berry curvature changes the instability criteria of the Landau's Fermi-liquid state. In particular, we reveal that even if the zero-sound mode is stable in the region of the forward-scattering amplitude, the Berry curvature causes Landau damping beyond the description of the Landau's Fermi-liquid theory. Based on the instability criteria of the zero-sound mode, we propose a phase diagram for a topological Fermi-liquid state against the phase separation, which generalizes the one-dimensional (Landau's interaction parameter) phase diagram of the Landau's Fermi-liquid state \cite{Negele_Orland,Nozieres_Pines} into a two-dimensional (Landau's interaction parameter and effective Berry curvature) one of a topological Fermi-liquid phase.

\section{Instabilities in a Landau's Fermi liquid state based on the Boltzmann transport theory}

In this section we review the Landau's Fermi-liquid theory based on the Boltzmann equation framework \cite{Negele_Orland,Nozieres_Pines}, applied and generalized into a topological Fermi-liquid theory \cite{TFL_Kim,Berry_FL}. We start from the Boltzmann transport equation
\begin{eqnarray}
	\frac{\partial n_{\bm{p}}(\bm{r},t)}{\partial t}+\dot{\bm{r}}\cdot\frac{\partial n_{\bm{p}}(\bm{r},t)}{\partial \bm{r}}+\dot{\bm{p}}\cdot\frac{\partial n_{\bm{p}}(\bm{r},t)}{\partial \bm{p}}=I[\delta{n}_{\bm{p}}(\bm{r},t)] .
\end{eqnarray}
Here, $n_{\bm{p}}(\bm{r},t)=n_{\bm{p}}^{0}+\delta n_{\bm{p}}(\bm{r},t)$ is the distribution function, where $n_{\bm{p}}^{0}$ is the equilibrium distribution function and $\delta n_{\bm{p}}(\bm{r},t)$ is its variation. During the evolution between successive collisions, the motion of electron quasiparticles on a Fermi surface obeys the Hamiltonian dynamics but with an additional interaction energy from other quasiparticles, described by
\begin{eqnarray}
	\dot{\bm{r}}&=&\frac{\partial\tilde{\epsilon}_{\bm{p}}(\bm{r},t)}{\partial \bm{p}} \\
	\dot{\bm{p}}&=&-\frac{\partial\tilde{\epsilon}_{\bm{p}}(\bm{r},t)}{\partial \bm{r}} .
\end{eqnarray}
Here, the quasiparticle energy $\tilde{\epsilon}_{\bm{p}}(\bm{r},t)$ is given by
\begin{eqnarray}
	\tilde{\epsilon}_{\bm{p}}(\bm{r},t)={\epsilon}^{0}_{\bm{p}}+\delta{\epsilon}_{\bm{p}}(\bm{r},t) ,
\end{eqnarray}
where the variation $\delta{\epsilon}_{\bm{p}}(\bm{r},t)$ results from renormalized forward scattering interactions, given by
\begin{eqnarray}
	\delta{\epsilon}_{\bm{p}}(\bm{r},t)=\sum_{\bm{p'}}f_{\bm{p}\bm{p'}}\delta n_{\bm{p'}}(\bm{r},t) .
\end{eqnarray}
$f_{\bm{p}\bm{p'}}$ is a Landau's parameter indicating the interaction strength of forward scattering between quasiparticles. It turns out that these effective interactions are marginal in the renormalization group analysis \cite{Shankar_RG_LFL}. In other words, quasiparticles are still interacting in the Landau's Fermi-liquid state, identifying such a state with an interacting fixed point instead of a non-interacting one. As a result, such effective interactions give rise to an effective potential field for a quasiparticle, described by $\delta{\epsilon}_{\bm{p}}(\bm{r},t)$. $I[\delta{n}_{\bm{p}}(\bm{r},t)]$ is an effective collision term. This collision integral can be approximated within the relaxation time approximation denoted by $\tau$. If one is interested in the regime of $\omega\gg\tau^{-1}$, which corresponds to the collisionless regime, the collision integral can be neglected.

Linearizing the Boltzmann equation with respect to $\delta{n}_{\bm{p}}(\bm{r},t)$ in the absence of external fields and performing the Fourier transformation, we obtain
\begin{eqnarray}
	&&(\omega-\bm{q}\cdot\bm{v}_{\bm{p}})\delta n_{\bm{p}}(\bm{q},\omega)\nonumber\\&&-\bm{q}\cdot\bm{v}_{\bm{p}}\left(-\frac{\partial n^{0}_{\bm{p}}}{\partial \epsilon_{p}}\right)\sum_{\bm{p'}}f_{\bm{p}\bm{p'}}\delta n_{\bm{p'}}(\bm{q},\omega)\nonumber\\&&=0 \label{Linearized_Boltzmann_Eq}
\end{eqnarray}

In order to solve this Boltzmann equation, we take an ansatz
\begin{eqnarray}
	\delta n_{\bm{p}}(\bm{q},\omega)=-\frac{\partial n^{0}_{\bm{p}}}{\partial \epsilon_{p}}\frac{\partial \epsilon_{p}}{\partial \bm{p}}\cdot\delta\bm{p}\equiv-\frac{\partial n^{0}_{\bm{p}}}{\partial \epsilon_{p}}v_{p}u({\theta,\phi}) .
\end{eqnarray}
Here, the group velocity $\bm{v}_{\bm{p}}=\frac{\partial \epsilon_{p}}{\partial \bm{p}}$ is parallel with $\delta\bm{p}$ near the Fermi surface. $u({\theta,\phi}) = |\delta\bm{p}|$ is an eigenvector, where our coordinate system assigns the direction of $\bm{q}$ with the $\bm{z}$-axis. Inserting this ansatz into Eq. (\ref{Linearized_Boltzmann_Eq}), we obtain
\begin{eqnarray}
	(\omega&-&\bm{q}\cdot\bm{v}_{\bm{p}})\left(-\frac{\partial n^{0}_{\bm{p}}}{\partial \epsilon_{p}}\right)v_{p}u({\theta,\phi})\nonumber\\
	&-&\bm{q}\cdot\bm{v}_{\bm{p}}\left(-\frac{\partial n^{0}_{\bm{p}}}{\partial \epsilon_{p}}\right)\sum_{\bm{p'}}f_{\bm{p}\bm{p'}}\left(-\frac{\partial n^{0}_{\bm{p'}}}{\partial \epsilon_{p'}}\right)v_{p'}u({\theta',\phi'})\nonumber\\
	&&=0 .
\end{eqnarray}

Performing the radial integration for the unprimed coordinate, we obtain
\begin{eqnarray}
	(\omega&-&{q}{v}_{F}\cos\theta)u({\theta,\phi})\nonumber\\
	&-&{q}{v}_{F}\cos\theta\sum_{\bm{p'}}f_{\bm{p}\bm{p'}}\left(-\frac{\partial n^{0}_{\bm{p'}}}{\partial \epsilon_{p'}}\right)u({\theta',\phi'})=0 ,
\end{eqnarray}
where we utilized $-\frac{\partial n^{0}_{\bm{p}}}{\partial \epsilon_{p}}\approx \delta(\epsilon_{p}-\epsilon_{F})$. Performing the radial integration also in the primed coordinate and introducing dimensionless parameters $s=\frac{\omega}{v_{F}q}$ and $F(\theta_{\bm{p}\bm{p'}})=\nu(\epsilon_{F})f_{\bm{p}\bm{p'}}$, where $\nu(\epsilon_{F})$ is the density of states at the Fermi surface, we reach the following expression
\begin{eqnarray}
	(s &-&\cos\theta)u\left(\theta,\phi\right)\nonumber\\
	&-&\cos\theta\int\frac{d\Omega'}{4\pi}F\left(\theta_{\bm{p}\bm{p'}}\right)u(\theta',\phi')=0 .
\end{eqnarray}
Here, $\int\frac{d\Omega'}{4\pi}$ means an angular integral for the momentum $\bm{p'}$.

Considering the spherical symmetry of our Fermi surface, we expand the eigenvector $u(\theta,\phi)$ and the interaction strength $F(\theta_{\bm{p}\bm{p'}})$ in the spherical harmonics and the Legendre polynomials, respectively, as follows
\begin{eqnarray}
	u(\theta,\phi)&=&\sum_{lm}Y_{lm}(\theta,\phi)u_{lm} , \\
	F(\theta_{\bm{p}\bm{p'}})&=&\sum_{l}P_{l}(\theta_{\bm{p}\bm{p'}})F_{l} .
\end{eqnarray}
As a result, we obtain
\begin{eqnarray}
	\sum_{lm}Y_{lm}\left(\theta,\phi\right)u_{lm}\left[\left(s-\cos\theta\right)-\cos\theta\frac{F_{l}}{2l+1}\right]=0 . \label{Dispersion_Collective_Modes_LFL}
\end{eqnarray}

An instability condition of this Fermi-liquid state is given by $s = 0$ and $u_{lm} \not= 0$, which means that Fermi-surface deformations $u_{lm}$ in the angular-momentum channel of $l$ with $m$ can occur without any energy cost given by $s = 0$. Here, we focus on the instability which preserves the translational symmetry. These instabilities arise when the dimensionless forward scattering parameter is given by \cite{Negele_Orland,Nozieres_Pines}
\begin{eqnarray}
	F_{l} \leq - (2l + 1) .
\end{eqnarray}

Given $F_{l}$, one can find the dispersion relation for Fermi-surface deformations in the angular momentum channel of $l$ and $m$. It is not straightforward to solve Eq. (\ref{Dispersion_Collective_Modes_LFL}) generally since various collective modes are correlated to result in coupled equations for Fermi-surface deformations. Here, we focus on the simplest case where only $F_{0}$ exists. Then, the calculation becomes straightforward to result in
\begin{eqnarray}
	\frac{s}{2}\ln\left(\frac{s+1}{s-1}\right)-1=\frac{1}{F_{0}}.
\end{eqnarray}
Solving this equation, we find the dispersion relation of the zero sound mode, which corresponds to the pole of the density-density correlation function in the Landau's Fermi-liquid state \cite{Negele_Orland,Nozieres_Pines}. This dispersion relation will be revisited below in comparison with that in a topological Fermi liquid state.

\section{Instabilities in a topological Fermi liquid state based on the Boltzmann equation framework}

\subsection{Boltzmann transport theory for a topological Fermi liquid state}

A minimal model for a Weyl metal phase with broken time reversal symmetry consists of a pair of chiral Fermi surfaces at a finite chemical potential \cite{WM1,WM2,WM3,WM4,WM_Review1,WM_Review2,WM_Review3}. Here, the term ``chiral" means that the chirality $+$ ($-$) is assigned to quasiparticles on one (the other) Fermi surface as a good quantum number. This Weyl band structure is realized, applying magnetic fields into a spin-orbit coupled Dirac metal state. Then, the Zeeman energy contribution splits the four-fold degeneracy into a pair of two-fold degeneracy, as discussed before. As a result, we write down our coupled Boltzmann equations as follows
\begin{eqnarray}
	&&\frac{\partial n^{\chi}_{\bm{p}}(\bm{r},t)}{\partial t}+\dot{\bm{r}}^{\chi}\cdot\frac{\partial n^{\chi}_{\bm{p}}(\bm{r},t)}{\partial \bm{r}}+\dot{\bm{p}}^{\chi}\cdot\frac{\partial n^{\chi}_{\bm{p}}(\bm{r},t)}{\partial \bm{p}}\nonumber\\
	&&=I[\delta n^{+}_{\bm{p}}(\bm{r},t), \delta n^{-}_{\bm{p}}(\bm{r},t)] .
\end{eqnarray}
Here, the superscript $\chi = \pm$ denotes the chirality of each Fermi surface. We emphasize that these coupled Boltzmann equations assume the limit of large effective spin-orbit interactions. This means that spin degrees of freedom are locked on the pair of chiral Fermi surfaces. As a result, we have effectively spinless fermions on the pair of chiral Fermi surfaces, where the role of spin degrees of freedom in scattering events appears as matrix elements for low-energy spinless chiral fermions. These effects are nothing but the Berry phase for spinless chiral fermions. In this respect an essential point is that the Hamiltonian dynamics of these chiral fermions are modified to incorporate such Berry-curvature effects. The pair of Weyl points given by two-fold degeneracy is mathematically described by a pair of magnetic monopoles in momentum space, responsible for the Berry curvature in the dynamics of chiral fermions \cite{WM_Review1,WM_Review2,WM_Review3}.

Their Hamiltonian dynamics are generalized in the presence of external electric ${\bm{E}}$ and magnetic $\bm{B}$ fields as follows \cite{TFL_Kim,Berry_FL,Anomalous_Velocity_Review1,Anomalous_Velocity_Review2,Boltzmann_Theory_WM1,Boltzmann_Theory_WM2,Boltzmann_Theory_WM3,Boltzmann_Theory_WM3_1,Boltzmann_Theory_WM4,Boltzmann_Theory_WM5,
Boltzmann_Theory_WM6,WM_Hamiltonian_Dynamics1,WM_Hamiltonian_Dynamics2,Disorder_WM}
\begin{eqnarray}
	G_{\chi}\dot{\bm{r}}^{\chi}&=&\left\{ \bm{v}^{\chi}_{\bm{p}}+e \tilde{\bm{E}}\times \bm{\Omega}^{\chi}_{\bm{p}} +\frac{e}{c}\big(\bm{\Omega}^{\chi}_{\bm{p}} \cdot \bm{v}^{\chi}_{\bm{p}}\big)\bm{B}\right\} , \\
	G_{\chi}\dot{\bm{p}}^{\chi}&=&\Big\{ e \tilde{\bm{E}} +\frac{e}{c}\big(\bm{v}^{\chi}_{\bm{p}}\times \bm{B}\big)+\frac{e^{2}}{c}\big(\tilde{\bm{E}} \cdot \bm{B} \big)\bm{\Omega}^{\chi}_{\bm{p}}\Big\} .
\end{eqnarray}
Here, $G_{\chi}=1+\frac{e}{c}\bm{B}\cdot\bm{\Omega}^{\chi}_{\bm{p}}$ is a modification factor for the phase-space volume, which originates from the magnetic-monopole singularity in the phase space. $\bm{\Omega}^{\chi}_{\bm{p}}$ is the Berry curvature, given by
\begin{equation}
\bm{\nabla}_{\bm{p}} \cdot \bm{\Omega}^{\chi}_{\bm{p}} = \chi \delta^{(3)}(\bm{p} - \bm{p}_{\chi}) ,
\end{equation}
where $\bm{p}_{\chi}$ is the position of the magnetic monopole in momentum space. $\bm{v}^{\chi}_{\bm{p}}=\frac{\partial\tilde{\epsilon}^{\chi}_{\bm{p}}}{\partial\bm{p}}$ is the group velocity of quasiparticle excitations on the chiral Fermi surface, where $\tilde{\epsilon}^{\chi}_{\bm{p}}=\epsilon^{\chi}_{\bm{p}}+\delta \epsilon^{\chi}_{\bm{p}}$ is the quasiparticle spectrum. $\epsilon^{\chi}_{\bm{p}}$ is the bare dispersion relation without interaction effects, given by
\begin{equation}
\epsilon^{\chi}_{\bm{p}}=\left(1-\frac{e}{c}\bm{B}\cdot\bm{\Omega}^{\chi}_{\bm{p}}\right)|\bm{p}| ,
\end{equation}
where the group velocity is renormalized to depend on the Berry curvature and to reflect the Lorentz invariance \cite{Boltzmann_Theory_WM1,Boltzmann_Theory_WM2,Boltzmann_Theory_WM3_1,Boltzmann_Theory_WM5,WM_Hamiltonian_Dynamics1,WM_Hamiltonian_Dynamics2}. The force equation should be modified due to interaction effects. Effective electric fields are given by \begin{equation}
\tilde{\bm{E}}=\bm{E}-\frac{1}{e}\frac{\partial\delta\epsilon^{\chi}_{\bm{p}}}{\partial\bm{r}} ,
\end{equation}
where
\begin{equation}
\delta\epsilon^{\chi}_{\bm{p}}=\sum_{\chi' = \pm}\sum_{\bm{p'}}f_{\bm{p}\bm{p'}}^{\chi\chi'} \delta n^{\chi'}_{\bm{p'}} \approx \sum_{p'}f_{\bm{p}\bm{p'}}\delta n^{\chi}_{\bm{p'}}
\end{equation}
is an interaction correction due to backflow contributions as that of the Landau's Fermi-liquid theory except for the presence of the pair of Fermi surfaces. Recently, one of the authors proved that these forward scattering interactions are marginal as those in the Landau's Fermi-liquid theory \cite{TFL_Kim}. This interacting fixed point with a pair of chiral Fermi surfaces was coined as a topological Fermi-liquid state and described by a topological Fermi-liquid theory, following Haldane's naming. In the present study we simplify these effective forward scattering interactions further. We consider interactions within the same chiral Fermi surface, where the chirality quantum number is preserved. In this respect one chiral Fermi surface does not communicate with the other. But, this zeroth-order approximation can be improved to take effective interactions between inter chiral Fermi surfaces.

Following the previous section, we focus on the collisionless regime to neglect the collision integral. Performing the Fourier transformation as follows
\begin{eqnarray}
	\delta n^{\chi}_{\bm{p}}(\bm{r},t)&=&\delta n^{\chi}_{\bm{p}}(\bm{q},\omega)e^{{i(\bm{q}\cdot\bm{r}-\omega t)}} , \nonumber \\
	\varphi&=&\varphi(\bm{q},\omega)e^{{i(\bm{q}\cdot\bm{r}-\omega t)}} , \nonumber \\
	\frac{\partial}{\partial \bm{r}}&\rightarrow& i\bm{q},~~~\frac{\partial}{\partial t}\rightarrow -i\omega , \nonumber
\end{eqnarray}
and linearizing the Boltzmann equation in $\delta n^{\chi}_{\bm{p}}(\bm{q},\omega)$ as before, we obtain
\begin{widetext}
\begin{eqnarray}
	&&\Big\{\omega-G^{-1}_{\chi}\left[\bm{v}^{\chi}_{\bm{p}}\cdot \bm{q}+\frac{e}{c}(\bm{\Omega}^{\chi}_{\bm{p}}\cdot\bm{v}^{\chi}_{\bm{p}})(\bm{B}\cdot\bm{q}) \right]\Big\}\delta n^{\chi}_{\bm{p}}(\bm{q},\omega) + G^{-1}_{\chi}\Big\{\bm{v}^{\chi}_{\bm{p}}\cdot \bm{q}+\frac{e}{c}(\bm{\Omega}^{\chi}_{\bm{p}}\cdot\bm{v}^{\chi}_{\bm{p}})(\bm{B}\cdot\bm{q}) \Big\} \frac{\partial n^{0}_{\chi}}{\partial \epsilon^{\chi}_{\bm{p}}}\sum_{\bm{p'}}f_{\bm{p}\bm{p'}}\delta n^{\chi}_{\bm{p'}}(\bm{q},\omega)\nonumber\\
	&+&e\varphi(\bm{q},\omega){G^{-1}_{\chi}\Big\{\bm{v}^{\chi}_{\bm{p}}\cdot \bm{q}+\frac{e}{c}(\bm{\Omega}^{\chi}_{\bm{p}}\cdot\bm{v}^{\chi}_{\bm{p}})(\bm{B}\cdot\bm{q}) \Big\}\frac{\partial n^{0}_{\chi}}{\partial \epsilon^{\chi}_{\bm{p}}}} = 0 .
\end{eqnarray}
\end{widetext}
Here, the longitudinal electric field is given by $\bm{E}=-\bm{\nabla}_{\bm{r}}\varphi(\bm{\bm{r}},t)$.

Turning off the external electric field, we obtain
\begin{eqnarray}
	&&\Big\{\omega-G^{-1}_{\chi}\left[\bm{v}^{\chi}_{\bm{p}}\cdot \bm{q}+\frac{e}{c}(\bm{\Omega}^{\chi}_{\bm{p}}\cdot\bm{v}^{\chi}_{\bm{p}})(\bm{B}\cdot\bm{q}) \right]\Big\}\delta n^{\chi}_{\bm{p}}(\bm{q},\omega)\nonumber\\
	&&+G^{-1}_{\chi}\Big\{\bm{v}^{\chi}_{\bm{p}}\cdot \bm{q}+\frac{e}{c}(\bm{\Omega}^{\chi}_{\bm{p}}\cdot\bm{v}^{\chi}_{\bm{p}})(\bm{B}\cdot\bm{q}) \Big\} \nonumber \\ &&\times \frac{\partial n^{0}_{\chi}}{\partial \epsilon^{\chi}_{\bm{p}}}\sum_{\bm{p'}}f_{\bm{p}\bm{p'}}\delta n^{\chi}_{\bm{p'}}(\bm{q},\omega) =0 . \label{Collective_Dynamics_TFLT}
\end{eqnarray}
Solving this equation is the main subject of this study.

\subsection{Instabilities in a topological Fermi liquid state}

Following the previous section, we introduce
\begin{eqnarray}
	\delta n_{\bm{p}}^{\chi}=-\frac{\partial n^{0}_{\chi}}{\partial\epsilon^{\chi}_{\bm{p}}}\bm{v}_{\bm{p}}^{\chi}\cdot\delta \bm{p} \label{Ansatz_TFLT}
\end{eqnarray}
in order to describe Fermi-surface deformations in the topological Fermi-liquid state. Since it is rather complex to consider a general situation, we focus on the case of $\bm{B}\parallel\hat{\bm{z}}\parallel\bm{q}$, i.e., $\bm{B}=B\hat{\bm{z}}$ and $\bm{q}=q\hat{\bm{z}}$, which corresponds to the case that the role of the Berry curvature is maximized. If the situation of $\bm{B}\cdot\bm{q} = 0$ is taken into account, Eq. (\ref{Collective_Dynamics_TFLT}) is reduced into the Landau's Fermi liquid theory.

In this ansatz, the group velocity is given by
\begin{eqnarray}
	\bm{v}_{\bm{p}}^{\chi}
%
%
	&=&\nabla_{\bm{p}}\left[\left(1-\frac{e}{c}\bm{\Omega}_{\bm{p}}^{\chi}\cdot \bm{B} \right)|\bm{p}| \right] \nonumber\\
    &=&\left(1+2\frac{e}{c}\bm{B}\cdot\bm{\Omega}_{\bm{p}}^{\chi} \right)\hat{\bm{p}}-\frac{e}{c}\left(\hat{\bm{p}}\cdot\bm{\Omega}_{\bm{p}}^{\chi} \right)\bm{B} , \label{Velocity_TFL}
\end{eqnarray}
and renormalized by the Berry curvature. Inserting Eq. (\ref{Velocity_TFL}) into Eq. (\ref{Ansatz_TFLT}), we obtain
\begin{eqnarray}
	\delta n_{\bm{p}}^{\chi}
%
%
	                                    &=&-\frac{\partial n^{0}_{\chi}}{\partial\epsilon^{\chi}_{\bm{p}}}\left\{\left(1+2\frac{e}{c}\bm{B}\cdot\bm{\Omega}_{\bm{p}}^{\chi} \right)\hat{\bm{p}}-\frac{e}{c}\left(\hat{\bm{p}}\cdot\bm{\Omega}_{\bm{p}}^{\chi}\right)\bm{B}\right\}\cdot\delta \bm{p} . \nonumber \\ \label{Ansatz_Further_Expression}
\end{eqnarray}

The forward-scattering part in Eq. (\ref{Collective_Dynamics_TFLT}) is
\begin{eqnarray}
	&& \int \frac{d^{3}p'}{(2\pi)^{3}} G_{\chi}f_{\bm{p}\bm{p'}}\left(-\frac{\partial n^{0}_{\chi}}{\partial\epsilon_{\bm{p'}}^{\chi}}\right)\bm{v}_{\bm{p'}}^{\chi}\cdot\delta \bm{p'}\nonumber \\ && = \int \frac{d^{3}p'}{(2\pi)^{3}} G_{\chi}f_{\bm{p}\bm{p'}}\left(-\frac{\partial n^{0}_{\chi}}{\partial\epsilon_{\bm{p'}}^{\chi}}\right)\nonumber\\
	&&\times\left\{\left(1+2\frac{e}{c}\bm{B}\cdot\bm{\Omega}_{\bm{p'}}^{\chi} \right)\hat{\bm{p'}}-\frac{e}{c}\left(\hat{\bm{p'}}\cdot\bm{\Omega}_{\bm{p'}}^{\chi}\right)\bm{B}\right\}\cdot\delta \bm{p'} .
\end{eqnarray}
Inserting
\begin{eqnarray}
	- \frac{\partial n^{0}_{\chi}}{\partial\epsilon_{\bm{p'}}^{\chi}} &\approx& \left(1-\chi\frac{e}{2c}\frac{\bm{B}\cdot\hat{\bm{p'}}}{\mu^2}\right)\delta\left(p'-\mu\right) \nonumber \\ &-& \chi\frac{e}{2c}\frac{\bm{B}\cdot\hat{\bm{p'}}}{\mu}\frac{d}{d p'}\delta\left(p'-\mu\right)
\end{eqnarray}
into the above expression, where $\mu$ is the chemical potential, we obtain
\begin{eqnarray}
	&& \int \frac{d^{3}p'}{(2\pi)^{3}} G_{\chi}f_{\bm{p}\bm{p'}}\left(-\frac{\partial n^{0}_{\chi}}{\partial\epsilon_{\bm{p'}}^{\chi}}\right)\bm{v}_{\bm{p'}}^{\chi}\cdot\delta \bm{p'}\nonumber\\
%
%
	&=&\int \frac{d^{3}p'}{(2\pi)^{3}} f_{\bm{p}\bm{p'}}
	\Bigg\{\delta\left(p'-\mu\right) \bm{\hat{p'}}\cdot\delta\bm{p'} \nonumber \\ &-& \bigg(\chi\frac{e}{2c}\frac{\bm{B}\cdot\hat{\bm{p'}}}{\mu}\frac{d}{dp'}\delta\left(p'-\mu\right)\bigg)\hat{\bm{p'}}\cdot\delta\bm{p'}\nonumber\\
	&+& \delta\left(p'-\mu\right)\Big[2\frac{e}{c}\left(\bm{B}\cdot\bm{\Omega}^{\chi}_{\bm{p'}}\right)\hat{\bm{p'}}\cdot\delta\bm{p'}\nonumber \\ &-& \frac{e}{c}\left(\hat{\bm{p'}}\cdot\bm{\Omega}^{\chi}_{\bm{p'}} \right)\left(\bm{B}\cdot\delta\bm{p'} \right)\Big] \Bigg\} . \label{Forward_Scattering_TFLT}
\end{eqnarray}

In order to describe collective dynamics of chiral Fermi-surface deformations, we expand the eigenvector $u(\theta_{\bm{p}},\phi_{\bm{p}})$ and the interaction strength $F(\theta_{\bm{p}\bm{p'}})$ in the spherical harmonics and the Legendre polynomials, respectively, as follows
\begin{eqnarray}
	\delta p &\equiv& u(\theta_{\bm{p}},\phi_{\bm{p}}) = \sum_{lm}Y_{lm}(\theta_{\bm{p}},\phi_{\bm{p}}) u_{lm} , \label{Delta_p_Spherical_Harmonics} \\
	F_{\bm{p}\bm{p'}}&\equiv&\frac{4\pi \mu^{2}f_{\bm{p}\bm{p'}}}{(2\pi)^{3}} = \sum_{l}F_{l}P_{l}(\theta_{\bm{p}\bm{p'}}) . \label{Interaction_Parameter_Legendre_Polynomials}
\end{eqnarray}
Here, we take further simplification that the collective dynamics does not depend on the azimuthal angle $\phi$, which fixes $m$ to be $m=0$, given by
\begin{eqnarray}
	\delta{p}=u(\theta_{\bm{p}})=\sum_{l}Y_{l0}(\theta_{\bm{p}})u_{l0} . \label{Delta_p_Spherical_Harmonics_Simplification}
\end{eqnarray}
This approximation has been performed in the Landau's Fermi liquid theory.

Inserting Eqs. (\ref{Ansatz_Further_Expression}) and (\ref{Forward_Scattering_TFLT}) into Eq. (\ref{Collective_Dynamics_TFLT}) and keeping the expression up to the first order in the applied magnetic field, we reformulate the resulting equation in terms of Eqs. (\ref{Delta_p_Spherical_Harmonics}) or (\ref{Delta_p_Spherical_Harmonics_Simplification}) and (\ref{Interaction_Parameter_Legendre_Polynomials}). During these calculations, we resort to the addition theorem \cite{Arfken_Weber}
\begin{eqnarray}
	P_{l'}(\cos\theta_{\bm{p}\bm{p'}})&=&\frac{4\pi}{2l'+1}\sum_{m'=-l'}^{l'}Y_{l'm'}(\Omega)Y^{*}_{l'm'}(\Omega')
%
%
\end{eqnarray}
with $Y^{*}_{l'm'}(\Omega') = (-1)^{m'} Y_{l'-m'}(\Omega')$ and the following identities of \cite{Arfken_Weber}
\begin{eqnarray}
	&& \int \frac{d\Omega_{\bm{p'}}}{4\pi}P_{l'}(\theta_{\bm{p}\bm{p'}})Y_{lm}(\theta_{\bm{p'}},\phi_{\bm{p'}}) \nonumber \\ && = \delta_{ll'}\frac{1}{2l'+1}Y_{l'm}(\theta_{\bm{p}},\phi_{\bm{p}})
\end{eqnarray}
and
\begin{eqnarray}
	&& \int d\Omega Y_{lm}(\Omega)Y_{l'm'}(\Omega)Y_{l''m''}(\Omega) \nonumber \\ && = \sqrt{\frac{(2l+1)(2l'+1)(2l''+1)}{4\pi}}\begin{pmatrix} l & l' &l'' \\m & m' & m''\end{pmatrix}\begin{pmatrix} l & l' &l'' \\0 & 0 & 0\end{pmatrix} . \nonumber \\
\end{eqnarray}
Here, $\begin{pmatrix} l & l' &l'' \\m & m' & m''\end{pmatrix}$ is the Wigner 3-j symbols \cite{Arfken_Weber}. An essential step is shown in appendix.

Performing the radial integration in the unprimed coordinate and both radial and angular integrations in the primed coordinate, we reach the following expression
\begin{widetext}
\begin{eqnarray}
	&&\frac{\mu^{2}}{(2\pi)^{3}}\sum_{l}\Bigg\{\Big[-\omega-3\omega\frac{e}{c} B\Omega_{F}\cos\theta+q\cos\theta + 4 q \frac{e}{c}B\Omega_{F}\cos^{2}\theta  \Big]u_{l0}Y_{l0}(\theta)\nonumber\\
	&& + \Big[q\cos\theta+3 q\frac{e}{c} B\Omega_{F}\cos^{2}\theta\Big]\frac{F_{l}}{2l+1}u_{l0}Y_{l0}(\theta)\nonumber\\
	&& + \Big[3 q\frac{e}{c}B\Omega_{F}\cos\theta\Big]u_{l0} \sum_{|l-1|\leq l'\leq l+1}\frac{F_{l'}}{2l'+1}Y_{l'0}(\theta)\Big[\sqrt{(2l+1)(2l'+1)}{(C_{ll'1})^{2}}\Big]\Bigg\} = 0 .
\end{eqnarray}
\end{widetext}
Here, $C_{l'l1}=\begin{pmatrix} l' & l &1 \\0 & 0 & 0\end{pmatrix}$ denotes the Wigner 3-j symbol and $\Omega_{F}$ is the Berry curvature at the Fermi surface.

Previously, we considered that the forward-scattering interaction does not depend on the Berry curvature as the zeroth-order approximation. However, there are matrix elements in the scattering amplitudes, which results from the spin-momentum locking. Actually, the forward-scattering amplitude has been proposed as
\begin{eqnarray}
	f_{\bm{p}\bm{p'}}&\rightarrow& f_{\bm{p}\bm{p'}}+\frac{e}{c}\bm{B}\cdot\left(\bm{\Omega}_{\bm{p}}+\bm{\Omega}_{\bm{p'}}\right)f^{B}_{\bm{p}\bm{p'}} ,
\end{eqnarray}
where $f^{B}_{\bm{p}\bm{p'}}$ is the scattering amplitude to depend on the Berry curvature \cite{Boltzmann_Theory_WM1}. Accordingly, we have
\begin{eqnarray}
	F_{l}&\rightarrow&F_{l}+\frac{e}{c}B\Omega_{F}\cos\theta F^{B}_{l} , \\
	3\frac{e}{c}B\Omega_{F}F_{l'}&\rightarrow&3\frac{e}{c}B\Omega_{F}\left(F_{l'}+\frac{1}{3}F^{B}_{l'}\right) .
\end{eqnarray}
As a result, we find our equation to describe instabilities in a topological Fermi-liquid state, given by
\begin{widetext}
\begin{eqnarray}
	&&\sum_{l}\Big\{-\omega-3\omega\frac{e}{c} B\Omega_{F}\cos\theta+q\cos\theta+4 q \frac{e}{c}B\Omega_{F}\cos^{2}\theta  \Big\}u_{l0}Y_{l0}(\theta)\nonumber\\
	&+&\sum_{l}\left\{\left[q\cos\theta+3 q\frac{e}{c} B\Omega_{F}\cos^{2}\theta\right]\frac{F_{l}}{2l+1}+\left[q\frac{e}{c}B\Omega_{F}\cos^{2}\theta\right]\frac{F^{B}_{l}}{2l+1}\right\}u_{l0}Y_{l0}(\theta)\nonumber\\
	&+&\sum_{l}\Big\{3 q\frac{e}{c}B\Omega_{F}\cos\theta\Big\}u_{l0}\sum_{|l-1|\leq l'\leq l+1}\frac{F_{l'}+\frac{1}{3}F^{B}_{l'}}{2l'+1}Y_{l'0}(\theta)\sqrt{(2l+1)(2l'+1)}{(C_{ll'1})^{2}} = 0 .
\end{eqnarray}
\end{widetext}

In order to find the instability condition, we set $\omega = 0$ and rewrite the above expression as follows
\begin{widetext}
\begin{eqnarray}
	&&\sum_{l}\cos\theta u_{l0}\Bigg\{\left(1+\frac{F_{l}}{2l+1}\right)Y_{l0}(\Omega) +\frac{e}{c}B\Omega_{\mu}\sum_{|l-1|\leq l'\leq l+1}\sqrt{(2l'+1)(2l+1)}(C_{1ll'})^{2} \nonumber \\ &&~~~~~~~~~~~~~~~~~~~\times\left[4+3\left(\frac{F_{l}}{2l+1}+\frac{F_{l'}}{2l'+1}\right) +\left(\frac{F^{B}_{l}}{2l+1}+\frac{F^{B}_{l'}}{2l'+1}\right)\right]Y_{l'0}(\Omega)\Bigg\} = 0 .
\end{eqnarray}
\end{widetext}
It is easy to read the instability condition given by
\begin{eqnarray}
F_{l} = - (2 l + 1) , ~~~~~ F^{B}_{l} = 2 l + 1 .
\end{eqnarray}
Although the instability criteria for $F_{l}$ look the same as those of the Landau's Fermi-liquid state, this occurs from the introduction of the Berry-curvature dependent interaction $F_{l}^{B}$. If we set $F_{l}^{B} = 0$, the instability criteria for $F_{l}$ depend on the external magnetic field or the effective Berry curvature. This will be clarified in the discussion of the zero-sound mode.

\subsection{Zero sound modes with $F_{0}$ and $F_{0}^{B}$ only}

Now, we investigate the dispersion relation of the zero sound mode in our topological Fermi-liquid state. Following the strategy of the Landau's Fermi-liquid theory, we consider only $F_{0}$ and $F_{0}^{B}$ forward scattering amplitudes. Then, we find the equation of motion for the zero sound mode
\begin{eqnarray}
	&&\left(s+3sb\cos\theta-\cos\theta-4b\cos^{2}\theta\right)u(\theta)\nonumber\\
%
%
	&&=\cos\theta F_{0}u_{00}Y_{00}+\sqrt{3}b\cos\theta \left(F_{0}+\frac{1}{3}F^{B}_{0}\right)u_{10}Y_{00}\nonumber\\
	&&~~~+3b\cos^{2}\theta \left(F_{0}+\frac{1}{3} F^{B}_{0}\right)u_{00}Y_{00} ,
\end{eqnarray}
where $s=\omega/q$ and $b=\frac{e}{c}B\Omega_{F}$. This expression is rewritten as
\begin{eqnarray}
	u(\theta)
%
%
	&=&\frac{F_{0}}{\sqrt{4\pi}}\frac{\cos\theta\left(u_{00}+\sqrt{3}b u_{10}\right) + 3b\cos^{2}\theta u_{00}}{s+\left(3s b-1\right)\cos\theta-4b\cos^{2}\theta} \nonumber \\ &+& \frac{b F^{B}_{0}}{\sqrt{4\pi}}\frac{\frac{1}{\sqrt{3}}\cos\theta u_{10}+\cos^{2}\theta u_{00}}{s+\left(3s b-1\right)\cos\theta-4b\cos^{2}\theta} ,
\end{eqnarray}
where $Y_{00}=\frac{1}{\sqrt{4\pi}}$.

Recalling $u(\theta)=\sum_{l}Y_{l0}(\theta)u_{l0}$, it is straightforward to find self-consistent equations for $u_{00}$ and $u_{10}$, given by
\begin{eqnarray}
	u_{00}&=&F_{0}\int^{1}_{-1}\frac{dx}{2}\frac{\left(u_{00}+\sqrt{3}b u_{10}\right)x+3b u_{00}x^{2}}{s+\left(3b s-1\right)x-4b x^{2}}\nonumber\\
	&&+b F^{B}_{0}\int^{1}_{-1}\frac{dx}{2}\frac{\frac{1}{\sqrt{3}}u_{10}x+u_{00}x^{2}}{s+\left(3b s-1\right)x-4b x^{2}} , \\
	u_{10}&=&F_{0}\int^{1}_{-1}\frac{dx}{2}\frac{\left(\sqrt{3}u_{00}+{3}b u_{10}\right)x^{2}+3\sqrt{3}b u_{00}x^{3}}{s+\left(3b s-1\right)x-4b x^{2}}\nonumber\\
	&&+b F^{B}_{0}\int^{1}_{-1}\frac{dx}{2}\frac{u_{10}x^{2}+\sqrt{3}u_{00}x^{3}}{s+\left(3b s-1\right)x-4b x^{2}} .
\end{eqnarray}

In order to simplify these self-consistent equations, we introduce
\begin{eqnarray}
	&& I_{n}(s;b) \equiv \int^{1}_{-1}\frac{dx}{2s}\frac{x^{n}}{1+\left(3b -1/s\right)x-4b x^{2}/s}\nonumber\\
	&& \approx \int^{1}_{-1}\frac{dx}{2}\left[\frac{x^{n}}{s-x}-\frac{3bs  x^{n+1}}{(s-x)^2}+\frac{4b x^{n+2}}{(s-x)^2}\right] ,
\end{eqnarray}
where the integral expression has been expanded up to the first order in the effective Berry curvature $b$. Then, we obtain
\begin{eqnarray}
	I_{1}(s;b) &=&(1-6bs)D(s)+\frac{bs}{s^{2}-1} , \\
	I_{2}(s;b) &=& s(1-7bs)D(s)+\frac{7}{3}b+\frac{b}{s^{2}-1} , \\
	I_{3}(s;b) &=& s^{2}(1-8bs)D(s)-\frac{1}{3}+\frac{8}{3}bs+\frac{bs}{s^{2}-1} ,
\end{eqnarray}
where we have
\begin{eqnarray}
	D(s) =\int^{1}_{-1}\frac{dx}{2}\frac{x}{s-x}=\frac{s}{2}\ln\left(\frac{s+1}{s-1}\right)-1 .
\end{eqnarray}

Based on these expressions, we rewrite the self-consistent equations as follows
\begin{eqnarray}
	&&u_{00} = F_{0}\left[\left(u_{00}+\sqrt{3}b u_{10}\right)I_{1}+3b u_{00}I_{2}\right]\nonumber\\
	&&+b F^{B}_{0}\left(\frac{1}{\sqrt{3}}u_{10}I_{1}+u_{00}I_{2}\right)\nonumber\\
	&&=\left[\left((1-3bs)D(s)+\frac{bs}{s^{2}-1}\right)F_{0}+bsD(s)F^{B}_{0}\right]u_{00}\nonumber\\
	&&+\sqrt{3}\left[bD(s)F_{0}+\frac{1}{3}bD(s)F^{B}_{0}\right]u_{10} ,
\end{eqnarray}
and
\begin{eqnarray}
	&&u_{10}=F_{0}\left(\left(\sqrt{3}u_{00}+{3}b u_{10}\right)I_{2}+3\sqrt{3}b u_{00}I_{3}\right)\nonumber\\
	&&+b F^{B}_{0}\left(u_{10}I_{2}+\sqrt{3}u_{00}I_{3}\right) \nonumber \\
	&&=\sqrt{3}\Bigg[\left(s(1-4bs)D(s)+\frac{4}{3}b+\frac{b}{s^{2}-1}\right)F_{0}\nonumber\\
	&&+\left(bs^{2}D(s)-\frac{b}{3}\right)F^{B}_{0}\Bigg]u_{00} \nonumber \\ && +3\left[bsD(s)F_{0}+\frac{1}{3}bsD(s)F^{B}_{0}\right]u_{10} .
\end{eqnarray}

The dispersion relation of the zero sound mode is given by the following secular equation
\begin{widetext}
\begin{eqnarray}
	\begin{vmatrix}
	\left((1-3bs)D(s)+\frac{bs}{s^{2}-1}\right)F_{0}+bsD(s)F^{B}_{0}-1 & \sqrt{3}bD(s)\left(F_{0}+\frac{1}{3}F^{B}_{0}\right)\\
	\sqrt{3}\left(s(1-4bs)D(s)+\frac{4}{3}b+\frac{b}{s^{2}-1}\right)F_{0}+\sqrt{3}\left(bs^{2}D(s)-\frac{b}{3}\right)F^{B}_{0} & 3bsD(s)\left(F_{0}+\frac{1}{3}F^{B}_{0}\right)-1
	\end{vmatrix}=0 .
\end{eqnarray}
\end{widetext}
Discarding terms proportional to $b^2$ explicitly, the secular equation leads to
\begin{eqnarray}
	\frac{s}{2}\ln\left(\frac{s+1}{s-1}\right)-1&=&\frac{1-bs\frac{F_{0}}{s^{2}-1}}{F_{0}+2bsF^{B}_{0}} , \label{ZS_TFL}
\end{eqnarray}
which determines the instability condition for the zero sound mode in our topological Fermi-liquid state.

In order to find the instability criteria, we expand the right hand side up to the first order of the dimensionless magnetic field $b$. Then, we obtain
\begin{eqnarray}
	\frac{s}{2}\ln\left(\frac{s+1}{s-1}\right)-1&=& \frac{1}{F_{0}} - b s \Big( \frac{1}{s^{2}-1} + \frac{2F^{B}_{0}}{F_{0}^{2}} \Big) . \label{Zero_Sound_Equation_TFL}
\end{eqnarray}

%
%

We consider
\begin{eqnarray}
	s = s_{0} + \delta s , \label{Zero_Sound_Expansion}
\end{eqnarray}
in the above equation, where $s_{0}$ satisfies the equation of the zero sound mode in the Landau's Fermi-liquid state, given by
\begin{eqnarray}
	\frac{s_{0}}{2}\ln\left(\frac{s_{0}+1}{s_{0}-1}\right)-1&=&\frac{1}{F_{0}} . \label{ZS_LFL_s0}
\end{eqnarray}
%
%
%
%
%
%
%
Inserting Eq. (\ref{Zero_Sound_Expansion}) into Eq. (\ref{Zero_Sound_Equation_TFL}) and keeping the resulting expression up to the linear order for $\delta s$, we find
\begin{eqnarray}
 \frac{\delta s}{s_{0}} = \frac{- b \left(\frac{1}{s^{2}_{0}-1}+\frac{2F^{B}_{0}}{F^{2}_{0}}\right)}{\frac{1}{s_{0}}\left(1+\frac{1}{F_{0}}\right)-\frac{s_{0}}{s^{2}_{0}-1}-b\left[\frac{2s^{2}_{0}}{(s^{2}_{0}-1)^{2}} -\left(\frac{1}{s^{2}_{0}-1}+\frac{2F^{B}_{0}}{F^{2}_{0}}\right)\right]} . \label{ZS_TFL_delta_s} \nonumber \\
\end{eqnarray}
%
%
%
%
%
Eq. (\ref{ZS_LFL_s0}) gives rise to $F_{0} = -1$ as the instability condition, which is nothing but that of the zero sound mode in the Landau's Fermi-liquid state \cite{Negele_Orland,Nozieres_Pines}. It is easy to see $s_{0} = 0$ when $F_{0} = -1$. Taking into account $\delta s / s_{0} \rightarrow 0$, we find
\begin{eqnarray}
 \frac{1}{s^{2}_{0}-1}+\frac{2F^{B}_{0}}{F^{2}_{0}} = 0 .
\end{eqnarray}
Inserting $s_{0} = 0$ into the above, we obtain $F_{0}^{B} = 1/2$. As a result, we find the instability condition of the zero sound mode in our topological Fermi-liquid state, given by $F_{0} = - 1$ and $F_{0}^{B} = 1/2$. But, we point out that the instability of the zero sound mode is driven by $F_{0}$, where $F_{0} = -1$ results in $\delta s = 0$ at the same time. The condition of $F_{0}^{B} = 1/2$ may be regarded as consistency for our approximation.

In order to investigate nature of the zero sound mode, we consider both real and imaginary parts as
\begin{eqnarray}
s=s_{1}-i s_{2} . \label{Real_Imaginary_ZS}
\end{eqnarray}
Inserting Eq. (\ref{Real_Imaginary_ZS}) into Eq. (\ref{ZS_TFL}), we obtain
\begin{widetext}
\begin{eqnarray}
	\begin{cases}
	\text{Re:}+\frac{s_{1}}{2}\ln\left[\frac{\sqrt{(s_{1}^{2}+s_{2}^{2}-1)^{2}+4s_{2}^2}}{(s_{1}-1)^{2}+s_{2}^2}\right] +\frac{s_{2}}{2}\arctan\left(\frac{2s_{2}}{s_{1}^{2}+s_{2}^{2}-1}\right)=1+\frac{F_{0}+b\left(2F^{B}_{0}s_{1}-F^{2}_{0} \frac{s_{1}(s^{2}_{1}+s^{2}_{2}-1)}{(s^{2}_{1}-s^{2}_{2}-1)^{2}+4s^{2}_{1}s^{2}_{2}}\right)}{F_{0}^{2}+4bF_{0}F^{B}_{0}s_{1}}  \\
	\text{Im:}-\frac{s_{2}}{2}\ln\left[\frac{\sqrt{(s_{1}^{2}+s_{2}^{2}-1)^{2}+4s_{2}^2}}{(s_{1}-1)^{2}+s_{2}^2}\right] + \frac{s_{1}}{2}\arctan\left(\frac{2s_{2}}{s_{1}^{2}+s_{2}^{2}-1}\right)=\frac{b\left(2F^{B}_{0}s_{2}-F^{2}_{0} \frac{s_{2}(s^{2}_{1}+s^{2}_{2}+1)}{(s^{2}_{1}-s^{2}_{2}-1)^{2}+4s^{2}_{1}s^{2}_{2}}\right)}{F_{0}^{2}+4bF_{0}F^{B}_{0}s_{1}}
	\end{cases} . \label{ZS_TFL_Real_Imaginary}
\end{eqnarray}
\end{widetext}
Neglecting $F_{0}^{B}$ in this expression, we obtain
\begin{widetext}
\begin{eqnarray}
	\begin{cases}
	\text{Re:}+\frac{s_{1}}{2}\ln\left[\frac{\sqrt{(s_{1}^{2}+s_{2}^{2}-1)^{2}+4s_{2}^2}}{(s_{1}-1)^{2}+s_{2}^2}\right] +\frac{s_{2}}{2}\arctan\left(\frac{2s_{2}}{s_{1}^{2}+s_{2}^{2}-1}\right)=1+\frac{1}{F_{0}}-b\frac{s_{1}(s_{1}^{2}+s_{2}^{2}-1)}{(s_{1}^{2}-s_{2}^{2}-1)^{2}+4s_{1}^{2}s_{2}^{2}} \\
	\text{Im:}-\frac{s_{2}}{2}\ln\left[\frac{\sqrt{(s_{1}^{2}+s_{2}^{2}-1)^{2}+4s_{2}^2}}{(s_{1}-1)^{2}+s_{2}^2}\right] +\frac{s_{1}}{2}\arctan\left(\frac{2s_{2}}{s_{1}^{2}+s_{2}^{2}-1}\right)=-b\frac{s_{2}(s_{1}^{2}+s_{2}^{2}+1)}{(s_{1}^{2}-s_{2}^{2}-1)^{2}+4s_{1}^{2}s_{2}^{2}}
	\end{cases} . \label{ZS_TFL_Real_Imaginary_FB_Zero}
\end{eqnarray}
\end{widetext}
Furthermore, this equation is reduced into that of the zero sound mode in the Landau's Fermi-liquid state, setting $b = 0$ as follows
\begin{widetext}
\begin{eqnarray}
	\begin{cases}
	\text{Re:}+\frac{s_{1}^{(0)}}{2}\ln\left[\frac{\sqrt{(s_{1}^{{(0)}2}+s_{2}^{{(0)}2}-1)^{2}+4s_{2}^{{(0)}2}}}{(s_{1}^{(0)}-1)^{2}+s_{2}^{{(0)}2}}\right] +\frac{s_{2}^{(0)}}{2}\arctan\left(\frac{2s_{2}^{(0)}}{s_{1}^{{(0)}2}+s_{2}^{{(0)}2}-1}\right)=1+\frac{1}{F_{0}} \\
	\text{Im:}-\frac{s_{2}^{(0)}}{2}\ln\left[\frac{\sqrt{(s_{1}^{{(0)}2}+s_{2}^{{(0)}2}-1)^{2}+4s_{2}^{{(0)}2}}}{(s_{1}^{(0)}-1)^{2}+s_{2}^{{(0)}2}}\right] +\frac{s_{1}^{(0)}}{2}\arctan\left(\frac{2s_{2}^{(0)}}{s_{1}^{{(0)}2}+s_{2}^{{(0)}2}-1}\right)= 0
	\end{cases} . \label{ZS_LFL_Real_Imaginary}
\end{eqnarray}
\end{widetext}

Considering $s_{1} = 0$ in Eq. (\ref{ZS_TFL_Real_Imaginary}), we obtain
\begin{eqnarray}
	\begin{cases}
	\text{Re:~}\frac{s_{2}}{2}\arctan\left(\frac{2s_{2}}{s^{2}_{2}-1}\right)=1+\frac{1}{F_{0}} \\
	\text{Im:~}bs_{2}\left(2F_{0}^{B}-F^{2}_{0}\frac{1}{s^{2}_{2}+1}\right)=0
	\end{cases} . \label{Instability_Condition_TFL_Imaginary}
\end{eqnarray}
These equations give rise to the same instability condition as Eqs. (\ref{ZS_LFL_s0}) and (\ref{ZS_TFL_delta_s}). In other words, $F_{0} = -1$ results in $s_{2} = 0$ and $F_{0}^{B} = 1/2$ is consistent with this condition.

Setting $b = 0$ in Eq. (\ref{Instability_Condition_TFL_Imaginary}), this equation is reduced into
\begin{eqnarray}
	\frac{s_{2}^{(0)}}{2}\arctan\left(\frac{2s_{2}^{(0)}}{s^{{(0)}2}_{2}-1}\right)=1+\frac{1}{F_{0}} . \label{Instability_Condition_LFL_Imaginary}
\end{eqnarray}
This equation is applicable to the case of $F_{0} < - 1$ in the Landau's Fermi-liquid state, where the real part turns out to vanish \cite{Negele_Orland,Nozieres_Pines}. Considering that the left hand side is an even function for $s_{2}$, we realize that this equation allows two solutions, one of which corresponds to an overdamped mode, but the other of which gives rise to an instability of the zero sound mode. This instability condition of the Landau's Fermi-liquid state turns out to be modified due to the presence of the effective Berry curvature term, denoted by $b$.

In order to investigate the role of the Berry curvature in the zero sound mode, we consider a perturbation approach in Eq. (\ref{ZS_TFL_Real_Imaginary_FB_Zero}), given by
\begin{eqnarray}
s_{1} = s_{1}^{(0)} + \delta s_{1} , ~~~~~ s_{2} = s_{2}^{(0)} + \delta s_{2} .
\end{eqnarray}
Here, $s_{1}^{(0)}$ and $s_{2}^{(0)}$ are the solution of Eq. (\ref{ZS_LFL_Real_Imaginary}), i.e., the Landau's Fermi-liquid state.

First, we focus on the case of $F_{0} < - 1$, where $s_{1}^{(0)} = 0$ and $s_{2}^{(0)}$ is given by Eq. (\ref{Instability_Condition_LFL_Imaginary}). Taking into account all terms up to the first order in the effective Berry curvature, we find
\begin{eqnarray}
	\begin{cases}
	\text{Re:~}\left\{-\frac{s_{2}^{(0)}}{s_{2}^{{(0)}2}+1}+\frac{1}{2}\arctan\left(\frac{2s_{2}^{(0)}}{s^{{(0)}2}_{2}-1}\right)\right\}\delta s_{2}^{(0)} = 0 \\
	\text{Im:~}\left\{-\frac{s_{2}^{(0)}}{s_{2}^{{(0)}2}+1}+\frac{1}{2}\arctan\left(\frac{2s_{2}^{(0)}}{s^{{(0)}2}_{2}-1}\right)\right\}\delta s_{1}^{(0)} =-b\frac{s_{2}^{(0)}}{s_{2}^{{(0)}2}+1}
	\end{cases} . \nonumber
\end{eqnarray}
Although the Berry curvature gives rise to a small value in the real part of the zero sound mode, the zero sound mode remains unstable.

Second, we consider the case of $F_{0} > 0$, where $s_{2}^{(0)} = 0$ and $s_{1}^{(0)}$ is given by Eq. (\ref{ZS_LFL_s0}). Here, we keep all terms up to the first order in the effective Berry curvature as before, given by
\begin{widetext}
\begin{eqnarray}
	\begin{cases}
	\text{Re:~}\left\{\frac{s_{1}^{(0)}}{s_{1}^{{(0)}2}-1} - \frac{1}{2}\ln\left(\frac{s_{1}^{(0)}+1}{s_{1}^{(0)}-1}\right) + b\frac{s_{1}^{{(0)}2}+1}{(s_{1}^{{(0)}2}-1)^{2}} \right\} \delta s_{1} = b\frac{s_{1}^{(0)}}{s_{1}^{{(0)}2}-1}  \\
	\text{Im:~}\left\{\frac{s_{1}^{(0)}}{s_{1}^{{(0)}2}-1} - \frac{1}{2}\ln\left(\frac{s_{1}^{(0)}+1}{s_{1}^{(0)}-1}\right) + b\frac{s_{1}^{{(0)}2}+1}{(s_{1}^{{(0)}2}-1)^{2}} \right\}\delta s_{2} = 0
	\end{cases} . \label{ZS_TFL_Delta}
\end{eqnarray}
\end{widetext}
The effective Berry curvature affects the real part of the zero sound mode. The imaginary part remains to be zero unless the following condition of
\begin{eqnarray}
 \frac{s_{1}^{(0)}}{s_{1}^{{(0)}2}-1} - \frac{1}{s_{1}^{(0)}} \Big( 1 + \frac{1}{F_{0}} \Big) + b \frac{s_{1}^{{(0)}2}+1}{(s_{1}^{{(0)}2}-1)^{2}}  = 0 \label{F0Large_Landau_Damping}
\end{eqnarray}
is satisfied. When this condition is fulfilled, the imaginary part becomes finite, which reflects Landau damping of the zero sound mode.

\begin{figure}
\begin{center}
  \vspace{0.cm}
  \begin{tabular}{cc}
    \includegraphics[scale=0.2]{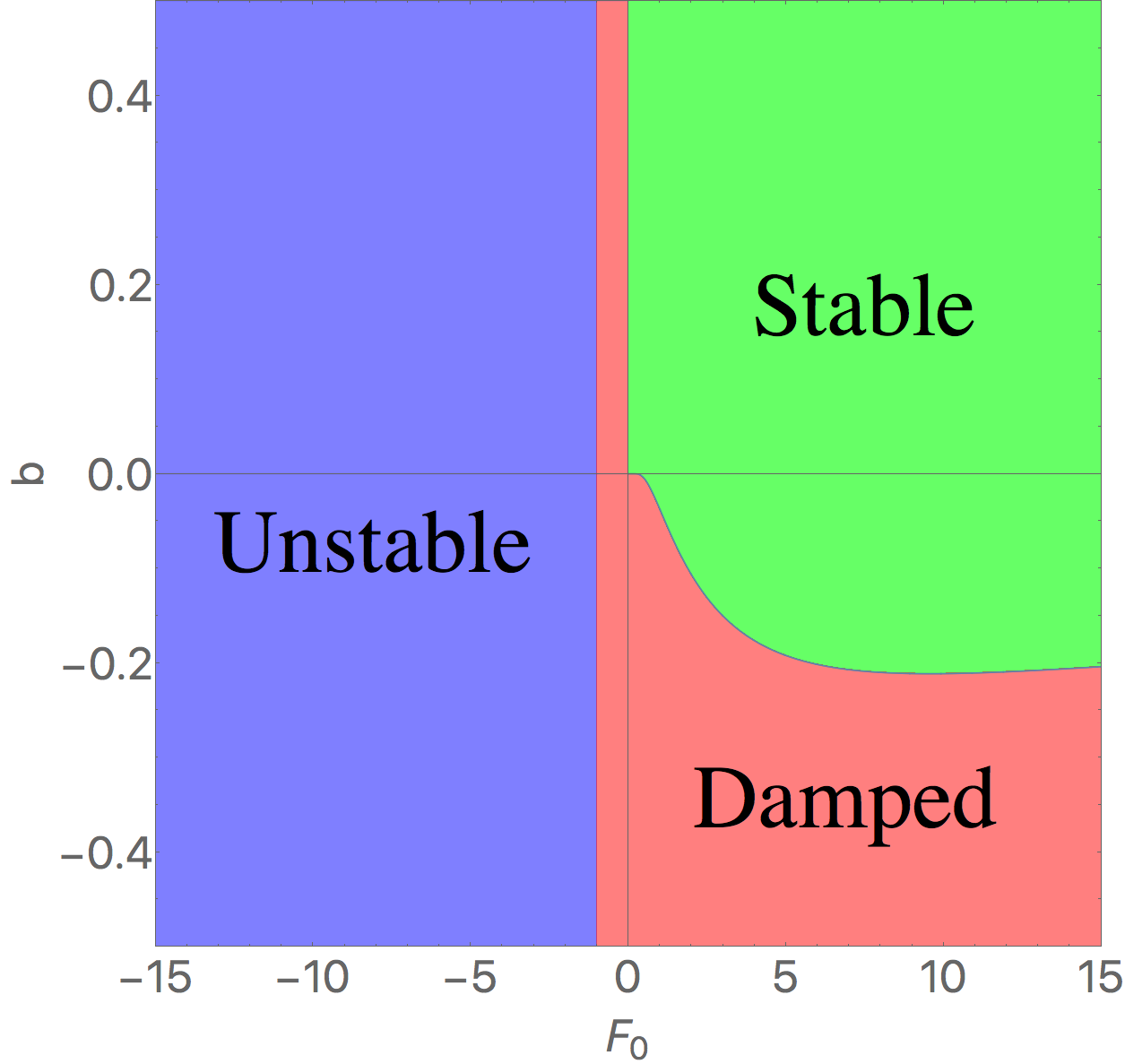}
    \end{tabular}
\end{center}
\vspace{-0.5cm}
\caption{Phase diagram based on the instability condition of the zero-sound mode in a topological Fermi-liquid state. Here, $F_{0}$ is the Landau's interaction parameter for forward scattering and $b$ is an effective Berry curvature. A $b = 0$ cut corresponds to the instability criterion of the zero-sound mode in a Landau's Fermi-liquid state. It turns out that the role of an effective Berry curvature $b$ changes the nature of the zero sound mode. An essential point beyond the Landau's Fermi-liquid phase is that the role of the Berry curvature gives rise to Landau damping even when $F_{0} > 0$.}
\vspace{-0.3cm}
\label{Phase_Diagram}
\end{figure}

%
%

The case of $- 1 < F_{0} \leq 0$ is quite complicated, where both real and imaginary parts are given by Eq. (\ref{ZS_LFL_Real_Imaginary}) \cite{Negele_Orland,Nozieres_Pines}. Repeating the above analysis, the small variation with respect to the solution of Eq. (\ref{ZS_LFL_Real_Imaginary}) up to the linear order of the effective Berry curvature is described by
\begin{widetext}
\begin{eqnarray}	 &&\Bigg\{\Bigg[\frac{-2s^{(0)}_{1}(s^{(0)2}_{1}+s^{(0)2}_{2}-1)+\left((s^{(0)}_{1}-1)^{2}+s^{(0)2}_{2}\right)\left((s^{(0)}_{1}+1)^{2}+s^{(0)2}_{2}\right)\ln\left[\frac{\sqrt{(s^{(0)2}_{1}+s^{(0)2}_{2}-1)^{2}+4s^{(0)2}_{2}}}{(s^{(0)}_{1}-1)^{2}+s^{(0)2}_{2}}\right]}{2\left((s^{(0)2}_{1}+s^{(0)2}_{2}-1)^{2}+4s^{(0)2}_{2}\right)}\Bigg]^{2}\nonumber\\
	&&+\Bigg[\frac{1}{2}\arctan\left(\frac{2s^{(0)}_{2}}{s^{(0)2}_{1}+s^{(0)2}_{2}-1}\right)-\frac{s^{(0)}_{2}(s^{(0)2}_{1}+s^{(0)2}_{2}+1)}{(s^{(0)2}_{1}+s^{(0)2}_{2})^{2}+4s^{(0)2}_{2}}\Bigg]^{2}\Bigg\}\delta s^{(0)}_{1}\nonumber\\ &&=\Bigg\{\frac{s^{(0)}_{1}(s^{(0)2}_{1}+s^{(0)2}_{2}-1)\ln\left[\frac{\sqrt{(s^{(0)2}_{1}+s^{(0)2}_{2}-1)^{2}+4s^{(0)2}_{2}}}{(s^{(0)}_{1}-1)^{2}+s^{(0)2}_{2}}\right]}{2\left((s^{(0)2}_{1}+s^{(0)2}_{2}-1)^{2}+4s^{(0)2}_{2}\right)}-\frac{s^{(0)}_{2}(s^{(0)2}_{1}+s^{(0)2}_{2}+1)\arctan\left(\frac{2s^{(0)}_{2}}{s^{(0)2}_{1}+s^{(0)2}_{2}-1}\right)}{2\left((s^{(0)2}_{1}+s^{(0)2}_{2}-1)^{2}+4s^{(0)2}_{2}\right)}\nonumber\\
	&&+\frac{(s^{(0)2}_{1}+s^{(0)2}_{2})}{(s^{(0)2}_{1}+s^{(0)2}_{2}-1)^{2}+4s^{(0)2}_{2}}\Bigg\}b
\end{eqnarray}
\end{widetext}
for the real part and
\begin{widetext}
\begin{eqnarray}
&&\Bigg\{\Bigg[\frac{-2s^{(0)}_{1}(s^{(0)2}_{1}+s^{(0)2}_{2}-1)+\left((s^{(0)}_{1}-1)^{2}+s^{(0)2}_{2}\right)\left((s^{(0)}_{1}+1)^{2}+s^{(0)2}_{2}\right)\ln\left[\frac{\sqrt{(s^{(0)2}_{1}+s^{(0)2}_{2}-1)^{2}+4s^{(0)2}_{2}}}{(s^{(0)}_{1}-1)^{2}+s^{(0)2}_{2}}\right]}{2\left((s^{(0)2}_{1}+s^{(0)2}_{2}-1)^{2}+4s^{(0)2}_{2}\right)}\Bigg]^{2}\nonumber\\
	&&+\Bigg[\frac{1}{2}\arctan\left(\frac{2s^{(0)}_{2}}{s^{(0)2}_{1}+s^{(0)2}_{2}-1}\right)-\frac{s^{(0)}_{2}(s^{(0)2}_{1}+s^{(0)2}_{2}+1)}{(s^{(0)2}_{1}+s^{(0)2}_{2})^{2}+4s^{(0)2}_{2}}\Bigg]^{2}\Bigg\}\delta s^{(0)}_{2}\nonumber\\ &&=\Bigg\{\frac{s^{(0)}_{2}(s^{(0)2}_{1}+s^{(0)2}_{2}+1)\ln\left[\frac{\sqrt{(s^{(0)2}_{1}+s^{(0)2}_{2}-1)^{2}+4s^{(0)2}_{2}}}{(s^{(0)}_{1}-1)^{2}+s^{(0)2}_{2}}\right]}{2\left((s^{(0)2}_{1}+s^{(0)2}_{2}-1)^{2}+4s^{(0)2}_{2}\right)}	 -\frac{s^{(0)}_{1}(s^{(0)2}_{1}+s^{(0)2}_{2}-1)\arctan\left(\frac{2s^{(0)}_{2}}{s^{(0)2}_{1}+s^{(0)2}_{2}-1}\right)}{2\left((s^{(0)2}_{1}+s^{(0)2}_{2}-1)^{2}+4s^{(0)2}_{2}\right)}\Bigg\}b
\end{eqnarray}
\end{widetext}
for the imaginary part. Both real and imaginary parts acquire $b-$linear corrections with respect to the Landau damping solution of Eq. (\ref{ZS_LFL_Real_Imaginary}).

The instability criteria for the zero sound mode suggest a phase diagram shown in Fig. \ref{Phase_Diagram}. Here, the phase diagram is extended from one dimension in a Landau's Fermi-liquid state to two dimensions in a topological Fermi-liquid phase, where $b$ is an effective Berry curvature and $F_{0}$ is an effective interaction parameter for forward scattering events. The one-dimensional line corresponding to a $b = 0$ cut serves as the instability criterion of the zero-sound mode in the Landau's Fermi-liquid state. It becomes destabilized when $F_{0} \leq - 1$, resulting in phase separation and given by the divergence of the compressibility \cite{Negele_Orland,Nozieres_Pines}. In this region of the interaction strength, the Berry curvature does not play a significant role in the instability of the zero sound mode. On the other hand, the role of the Berry curvature changes the nature of the zero sound mode drastically in the case of $F_{0} > 0$. In particular, the effective Berry curvature gives rise to Landau damping for the zero sound mode beyond the description of the Landau's Fermi-liquid theory. The boundary between undamped and Landau damped zero sound modes is determined by Eq. (\ref{F0Large_Landau_Damping}) in the case of $F_{0} > 0$. This Landau damped dynamics of the zero sound mode is smoothly connected with that in the region of $- 1 < F_{0} \leq 0$ with $b = 0$, i.e., the zero sound mode of the Landau's Fermi liquid state.

We would like to point out that only one chiral Fermi surface has been taken into account until now. More precisely, the pair of chiral Fermi surfaces are assumed to be independent as the zeroth-order approximation. Here, one chiral Fermi surface is characterized by a positive Berry curvature $b > 0$ while the other is identified with a negative one $b < 0$. In this respect, even if the zero-sound mode is undamped for one chiral Fermi surface, it can be Landau damped for the other one. Since we are considering that these chiral Fermi surfaces do not communicate with each other through Fermi-liquid interactions, we conclude that the undamped zero sound mode in one chiral Fermi surface coexists with the Landau damped one in the other chiral Fermi surface.

\section{Summary}

In summary, we investigated how an interacting Weyl metal phase with broken time reversal symmetry becomes destabilized when both the Berry curvature and chiral anomaly are introduced into a Landau's Fermi-liquid state. Based on the Boltzmann equation framework and following the Landau's Fermi-liquid theory, we derived eigenvalue problems, where eigenvectors describe Fermi-surface deformations and eigenvalues represent dispersion relations of collective dynamics of Fermi-surface fluctuations. Solving these coupled equations, where the coupling occurs between different angular momentum channels, we found that the role of the Berry curvature modifies the instability criteria of the Landau's Fermi-liquid state. In order to clarify this modification, we examined the zero sound mode for more details, described by the zero angular-momentum channel in our eigenvalue problems, where two parameters of the forward-scattering interaction $F_{0}$ and the effective Berry curvature $b$ or the applied magnetic field appear to control the dynamics of the zero sound mode. Our main result was that even if the zero sound mode is undamped due to the interaction effect, the role of the Berry curvature leads it to be Landau damped. This magnetic-field control for the collective dynamics of Fermi-surface fluctuations is beyond the Landau's Fermi-liquid state, regarded to be a characteristic feature of an interacting Weyl metal phase.

\section*{ACKNOWLEDGEMENT}

This study was supported by the Ministry of Education, Science, and Technology (No. NRF-2015R1C1A1A01051629 and No. 2011-0030046) of the National Research Foundation of Korea (NRF).

\appendix*

\section{How to represent the interaction part of the Boltzmann equation in terms of spherical harmonics}

In this appendix, we show how to decompose the interaction part of the Boltzmann equation in terms of the angular momentum. The interaction part is calculated as follows
\begin{widetext}
\begin{eqnarray}
	&&\int \frac{d^{3}p'}{(2\pi)^{3}} G_{\chi}f_{\bm{p}\bm{p'}}\left(-\frac{\partial n^{0}_{\chi}}{\partial\epsilon_{p'}^{\chi}}\right)\bm{v}_{\bm{p'}}^{\chi}\cdot\delta \bm{p'}\nonumber\\
	&=&\sum_{ll'm}\int\frac{p'^{2}dp'd(\cos\theta')d\phi'}{(2\pi)^{3}}f_{l'}P_{l'}(\cos\theta_{\bm{p}\bm{p'}})Y_{lm}(\theta',\phi')u_{lm}\nonumber\\
	&&~~~~\times\left\{\delta(p'-\mu)+\cos\theta'\left[-\frac{eB}{2c}\frac{1}{\mu}\frac{d}{dp'}\delta(p'-\mu)+\frac{eB}{2c}\frac{1}{\mu^2}\delta(p'-\mu)\right]\right\}\\
	&=&\sum_{ll'm}\int\frac{d\Omega'}{(2\pi)^3}\mu^{2}f_{l'}u_{lm}\left[P_{l'}(\cos\theta_{\bm{p}\bm{p'}})Y_{lm}(\Omega')+\frac{eB}{c}\frac{3}{2\mu^2}\cos\theta'P_{l'}(\cos\theta_{\bm{p}\bm{p'}})Y_{lm}(\Omega')\right]\\
	&=&\sum_{ll'm}F_{l'}u_{lm}\left[\delta_{ll'}\frac{1}{2l'+1}Y_{l'm}(\Omega)+\frac{eB}{c}\frac{3}{2\mu^2}\int\frac{d\Omega'}{4\pi}P_{l'}(\cos\theta_{\bm{p}\bm{p'}})Y_{lm}(\Omega')\cos\theta'\right]\\
	&=&\sum_{ll'm}F_{l'}u_{lm}\left[\delta_{ll'}\frac{1}{2l'+1}Y_{l'm}(\Omega)+\frac{eB}{c}\frac{3}{2\mu^2}\frac{1}{2l'+1}\sum_{m'=-l'}^{l'}Y_{l'm'}(\Omega)\sqrt{\frac{4\pi}{3}}\int d\Omega'Y^{*}_{l'm'}(\Omega')Y_{lm}(\Omega')Y_{10}(\Omega')\right] \\
	&=&\sum_{ll'm}F_{l'}u_{lm}\Bigg[\delta_{ll'}\frac{1}{2l'+1}Y_{l'm}(\Omega)+\frac{eB}{c}\frac{3}{2\mu^2}\frac{1}{2l'+1}\nonumber\\
	&&~~~~~~~~~~~~~~~~~\times\sum_{m'=-l'}^{l'}Y_{l'm'}(\Omega)\sqrt{(2l'+1)(2l+1)}\begin{pmatrix} l' & l &1 \\-m' & m & 0\end{pmatrix}\begin{pmatrix} l' & l &1 \\0 & 0 & 0\end{pmatrix}(-1)^{m'}\Bigg]\\
	&=&\sum_{ll'}F_{l'}u_{l0}\Bigg[\delta_{ll'}\frac{1}{2l'+1}Y_{l'0}(\theta)+\frac{eB}{c}\frac{3}{2\mu^2}\frac{1}{2l'+1}Y_{l'0}(\theta)\sqrt{(2l'+1)(2l+1)}(C_{l'l1})^{2}\Bigg] ,
\end{eqnarray}
\end{widetext}
where $C_{l'l1}=\begin{pmatrix} l' & l &1 \\0 & 0 & 0\end{pmatrix}$ is the Wigner 3-j symbol, given by $(C_{l'l1})^{2}=\frac{(-1)^{2l}(l+l^{2}-l'-l'^{2})^{2}}{(1+l-l')!(1-l+l')!(l+l')(1+l+l')(2+l+l')}$ \cite{Arfken_Weber}.
In the last equality, the independence of the azimuthal angle was assumed.

\end{document}